\documentstyle[aps,preprint]{revtex}

\begin{document}
\draft
\tightenlines
\title
{\bf Ginzburg-Landau Expansion and the Slope of the Upper Critical Field
in Superconductors with Anisotropic Normal Impurity Scattering}
\author{A.I.Posazhennikova,\ M.V.Sadovskii}
\address
{Institute for Electrophysics,\\
Russian Academy of Sciences,\ Ural Branch, \\ Ekaterinburg,\ 620049, Russia\\
E-mail:\ sadovski@ief.intec.ru} 
\maketitle

\begin{center}
{\sl Submitted to JETP,\ May 1997}
\end{center}

\begin{abstract}
Ginzburg-Landau expansion for superconductors with anisotropic $s-$ and
$d-$wave pairing is derived in the presence of anisotropic normal impurity 
scattering which makes $d-$pairing state more stable under disordering.
It is demonstrated that the slope of the upper critical field  
$|dH_{c2}/dT|_{T_{c}}$ in superconductors with $d-$wave pairing has nonlinear 
dependence on disorder, i.e. for the low anisotropic scattering rate it drops 
rather fast with concentration of normal impurities, but as anisotropy of
scattering 
increases it features initial nonlinear growth, approaches a maximum and 
drops again, vanishing at the critical impurity 
concentration. In superconductors with anisotropic $s-$wave pairing   
$|dH_{c2}/dT|_{T_{c}}$ grows, approaching the known asymptotic behavior, 
characteristic of usual isotropic case irrespective to the scattering
anisotropy.
\end{abstract}

\newpage
\narrowtext
\section{Introduction}
The main problem of the present day physics of high-temperature 
superconductors is the  determination of the nature and type of Cooper 
pairing. A majority of experiments and theoretical models \cite{DP} suggest the 
realization in these systems of anisotropic pairing with 
$d_{x^2-y^2}$-symmetry with zeroes of the gap function at the 
Fermi surface. At the same time there exist some theoretical and experimental 
evidences\cite{CS,LM} supporting the so-called anisotropic $s-$wave pairing. 
In this latest 
case there again appear zeroes (with no change of sign) or rather deep minima 
of the gap function in the same directions in the Brillouin zone as in the case 
of $d-$wave pairing.
It was shown \cite{BH,FN} that controlled disordering (introduction of normal 
impurities) can be an effective method of experimental discrimination between 
different types of anisotropic pairing. In our previous paper it was shown
that one can use measurements of the slope of the upper critical field
$|dH_{c2}/dT|_{T_{c}}$ for the same purpose, i.e. for the case of $d-$wave 
pairing the slope drops rather fast with disorder, while for the case  of 
anisotropic $s-$wave pairing it grows with disorder as for the isotropic 
case. 

Recently an interesting theoretical model was proposed\cite{HN} taking into
account the effects of anisotropic 
(momentum-dependent) impurity scattering. It was shown that for large enough 
anisotropic "$d-$wave" scattering the usual pair-breaking effect of
normal impurities (described by Abrikosov-Gorkov dependence for isotropic 
superconductor with magnetic impurities) is rather strongly suppressed. 
This allows, in principle at least, to overcome one of the main
problems in the physics of high-temperature superconductors --- the
contradiction between the $d-$wave picture of pairing in these systems and
their relative stability towards disordering\cite{S}. 
There exist also some other explanations of such a stability 
(cf. an explanation proposed in our paper \cite{SP}), 
however the simplicity 
of the suggested model \cite{HN} is rather attractive and calls for further 
theoretical study of superconductors with exotic pairing with the 
account of anisotropic impurity scattering.
The present paper is a natural continuation of our previous study, including 
the effects of anisotropic impurity scattering. It is demonstrated that
anisotropic impurity scattering leads to significant anomalies in 
the dependence of the slope of the upper critical field on disorder (impurity 
concentration). As in the previous paper \cite{PS} our analysis is based 
on the microscopic derivation of Ginzburg-Landau  expansion in impure system.
  
\section{Ginzburg-Landau expansion}
Following Refs.\cite{BH,FN}, we analyze two-dimensional electronic system 
with isotropic Fermi surface and separable pairing potential of the form:

\begin{equation} 
V({\bf p,p'})\equiv V(\phi ,\phi')=-Ve(\phi)e(\phi'), 
\label{1} 
\end{equation} 
where $\phi$ is a polar angle, determining the electronic momentum direction 
in the conducting plane, and $e(\phi)$ is given by the following model 
dependence
\begin{equation} 
e(\phi)=\left\{\begin{array}{ll} 
\sqrt{2}cos\left(2\phi\right) & ($d$-\mbox{ wave }), 
\\ \sqrt{2}|cos\left(2\phi\right)| & (\mbox{ anisotropic } 
$s$-\mbox{ wave }). 
\end{array} \right.  
\label{2} 
\end{equation}

The pairing constant $V$ is as usual different from zero in some region of 
the width of $2\omega_{c}$ around the Fermi level ($\omega_{c}$-is some 
characteristic frequency of the quanta, responsible for the pairing 
interaction). In this case the superconducting gap (order parameter) takes 
the form:
\begin{equation}
\Delta({\bf p})\equiv\Delta(\phi)=\Delta e(\phi), 
\label{3}
\end{equation}
and positions of its zeroes for $s$ and $d$ cases just coincide.

We examine a superconductor containing "normal" (nonmagnetic) impurities, 
which are chaotically distributed in space with concentration $\rho$.  
Following Ref.\cite{HN} we consider the square of the scattering
amplitude of the impurity in the following form:  
\begin{equation} 
|V_{imp}({\bf p,p'})|^2\equiv|V_{imp}(\phi,\phi')|^2 
=|V_0|^2+|V_1|^2f(\phi)f(\phi') , \label{4} 
\end{equation} 
where $V_0$ is 
isotropic point-like scattering amplitude, $V_1$ is anisotropic scattering 
amplitude,and $f(\phi)$ is angular-dependent model function ($\phi$ is a 
polar angle mentioned above) which describes the type of anisotropic 
scattering. We consider the scattering to be essentially isotropic and impose 
the following constraints \cite{HN}:  
\begin{equation} 
|V_1|^2\leq|V_0|^2;\ <f>=0;\  
<f^2>=1, \label{5} 
\end{equation} 
where $<...>$ denotes the average value 
over the momentum direction on the Fermi surface (i.e. over the 
$\phi$-angle).Accordingly, the second part in Eq.(\ref{4}) represents the 
deviations from the isotropic scattering.

The normal and anomalous temperature Green's functions in such a 
superconductor are\cite{AG}:
\begin{eqnarray}
G(\omega,{\bf p})=-\frac{i\tilde\omega+\xi_{\bf p}}{\tilde\omega^2+
\xi_{\bf p}^2+|\tilde\Delta({\bf p})|^2}, \label{6} \\
F(\omega,{\bf p})=\frac{\tilde\Delta^*({\bf p})}
{\tilde\omega^2+\xi_{\bf p}^2+ |\tilde\Delta({\bf p})|^2}, \nonumber
\end{eqnarray}
where $\omega=(2n+1)\pi T$, $\xi-$ is electronic energy with respect to the 
Fermi level,
\begin{eqnarray} 
\tilde\omega({\bf p})
=\omega+i\rho\int\frac{d{\bf p'}}{(2\pi)^2} |V_{imp}({\bf p-p'})|^2
G(\omega,{\bf p'}), \label{7} \\ 
\tilde\Delta({\bf p})=\Delta({\bf 
p})+\rho\int\frac{d{\bf p'}}{(2\pi)^2} |V_{imp}({\bf p-p'})|^2F(\omega,
{\bf p'}). 
\nonumber 
\end{eqnarray}

To determine the transition temperature we can limit ourselves to
Eqs.(\ref{7}) linearized over $\Delta$:  
\begin{eqnarray} 
\tilde\omega=\omega+i\rho\frac{N(0)}{2\pi} 
\int{d\xi}\int_{0}^{2\pi}{d\phi}
\left\{|V_0|^2+
|V_1|^2 f(\phi)f(\phi')\right\}\frac{\tilde\omega}{\tilde\omega^2+\xi^2},
\label{8}\\
\tilde\Delta=\Delta+\rho\frac{N(0)}{2\pi}
\int{d\xi}\int_{0}^{2\pi}{d\phi}
\left\{|V_0|^2+
|V_1|^2 f(\phi)f(\phi')\right\}\frac{\tilde\Delta}{\tilde\omega^2+\xi^2}.
\nonumber
\end{eqnarray}

The critical temperature $T_{c}$ is determined by the linearized 
gap-equation: 
\begin{equation}
\Delta({\bf p})=-T_{c}\sum_{\omega}\int\frac{d{\bf p'}}{(2\pi)^2}
V({\bf p,p'})\frac{\tilde\Delta({\bf p'})} {\tilde\omega^2+ \xi_{\bf p'}^2} 
\label{9} 
\end{equation} 
Following standard procedure from  Eqs.(\ref{8}),(\ref{9})
we obtain the following general equation for the 
critical temperature $T_{c}$: 
\begin{equation}
ln\biggl(\frac{T_{c}}{T_{c0}}\biggr)=\biggl( <e>^2+<ef>^2-1\biggr)
\biggl[\Psi\biggl(\frac{1}{2}+\frac{\gamma_0}{2\pi T_{c}}\biggr)-
\Psi\biggl(\frac{1}{2}\biggr)\biggr]+
\end{equation}
$$<ef>^2
\biggl[\Psi\biggl(\frac{1}{2}\biggr)-
\Psi\biggl(\frac{1}{2}+\frac{\gamma_0}{2\pi T_{c}}
\biggl(1-\frac{\gamma_{1}}{\gamma_{0}}\biggr)\biggr)
\biggr] $$
where $T_{c0}$- is the transition temperature in the absence of impurities, 
$\Psi(x)$- is the usual digamma function, $\gamma_{0}=\pi\rho V_0^2N(0)$ and 
$\gamma_{1}=\pi\rho V_1^2N(0)$ - correspondingly the isotropic and 
anisotropic impurity scattering rates, $<ef>^2$ describes the ``overlap'' 
between the functions $e(\bf p)$ and $f(\bf p)$. 

For the sake of simplicity we take the function $f(\bf p)$ in the form 
analogous to that in Eq.(\ref{2}):
\begin{equation} 
f({\bf p})\equiv f(\phi)=\sqrt{2}cos(2\phi), 
\label{.} 
\end{equation}
This corresponds to the maximum overlap for $d$-case. More general treatment
one could find in Ref.\cite{HN}. Now the renormalized Eqs.(\ref{8}) can be 
written as follows:  
\begin{eqnarray} 
\tilde\omega=\omega+i\frac{\gamma_0}{\pi}\int{d\xi}
\frac{\tilde\omega}{\tilde\omega^2+\xi^2}+i\frac{\gamma_1}{\pi^2}
cos(2\phi)\int{d\xi}\int{d\phi'}cos(2\phi')
\frac{\tilde\omega}{\tilde\omega^2+\xi^2},
\label{7a} \\ 
\tilde\Delta=\Delta+i\frac{\gamma_0}{\pi}\int{d\xi}
\frac{\tilde\Delta}{\tilde\omega^2+\xi^2}+i\frac{\gamma_1}{\pi^2}
cos(2\phi)\int{d\xi}\int{d\phi'}cos(2\phi')
\frac{\tilde\Delta}{\tilde\omega^2+\xi^2}. \nonumber
\end{eqnarray}
From here we obtain the well-known expression for the renormalized frequency 
in both cases:  
\begin{equation} \tilde\omega=\omega+\gamma_0 sign\omega.  
\label{13a}
\end{equation}

In the case of $d-$wave pairing the gap symmetry in the presence of 
impurities is not changed:  
\begin{equation} 
\tilde\Delta=\Delta\frac{|\tilde\omega|}{|\tilde\omega|-\gamma_{1}}.
\label{14}
\end{equation}
In the case of $s-$wave pairing the gap is shifted by a constant, 
which does not depend on $\phi$ and $\gamma_{1}$:  
\begin{equation} 
\tilde\Delta=\Delta+\Delta_{0}\frac{2\sqrt{2}\gamma_{0}}{\pi|\omega|}.
\label{15}
\end{equation}
Finally $T_{c}$-equation for superconductor with $d-$wave pairing is 
written as:
\begin{equation}
ln\biggl(\frac{T_{c}}{T_{c0}}\biggr)=
\Psi\biggl(\frac{1}{2}\biggr)-
\Psi\biggl(\frac{1}{2}+\biggl(1-\frac{\gamma_{1}}{\gamma_{0}}\biggr)
\frac{\gamma_0}{2\pi T_{c}}\biggr).
\label{16}
\end{equation}
For superconductor with anisotropic $s-$wave pairing $T_c$-equation is 
written as:
\begin{equation}
ln\biggl(\frac{T_{c}}{T_{c0}}\biggr)=\left(1-\frac{8}{\pi^2}\right)
\left[\Psi\biggl(\frac{1}{2}+\frac{\gamma_0}{2\pi T_{c}}\biggr)-
\Psi\biggl(\frac{1}{2}\biggr)\right].
\label{17}
\end{equation}
Note that anisotropic scattering rate dependence drops from Eq.(\ref{17}).

The appropriate dependencies of $T_{c}(\gamma_{0}/T_{c0})$ are shown in Fig.1, 
for the case of $d-$wave pairing with different values of the normalized 
anisotropic scattering rate  $\gamma_{1}/\gamma_{0}$.\ 
In the case of $s-$wave pairing the transition temperature $T_{c}$ is 
slightly suppressed with the growth of $\gamma_{0}/T_{c0}$. In the case of 
$d-$wave pairing $T_{c}$ suppression is rather fast for small values of 
$\gamma_{1}$, but the critical value of $\gamma_{0c}/T_{c0}$ leading to the 
complete destruction of superconducting state rapidly increases with the 
growth of the anisotropic scattering rate $\gamma_{1}/\gamma_{0}$.

The gap function as usual can be used as an order parameter
in the Ginzburg-Landau expansion  for the free-energy density. 
We consider 
its amplitude $\Delta(T)$ to be a slowly varying function of the spatial 
coordinates.  Accordingly in momentum space we get the Fourier-component of 
the order parameter:  
\begin{equation} \Delta(\phi,q)=\Delta_q(T)e(\phi).  
\label{18}
\end{equation}

The Ginzburg-Landau expansion for the free energy density difference between
superconducting and normal state in the region of small $q$ up to terms 
quadratic over $\Delta$ takes the form:
\begin{equation}
F_{s}-F_{n}=A|\Delta_{q}|^2+q^2 C|\Delta_{q}|^2 \label{6a}
\end{equation}
and is determined by the diagrams of loop-expansion for the free-energy of  
electrons moving in the field of superconducting order parameter fluctuations 
with some small vector $q$, shown in Fig.2. Diagrams (c) and (d) are to be 
subtracted, so that the coefficient $A$ becomes zero at the transition point  
$T=T_{c}$. 

Some details on calculating $\Gamma_{\bf pp'}$  and Ginzburg-Landau 
coefficients for the case of $d-$wave pairing can be found in the 
Appendices A and B. Note, that for the $d-$wave superconductors the 
contribution of diagrams Fig.2(b,d) actually vanishes up to 
terms of the 
order of $q^2$, if we don't take into account an anisotropy of impurity 
scattering. In the case of $s-$wave superconductor all calculations are 
analogous, we only note that in such a case a dependence on anisotropic 
scattering rate is absent.  

Finally we can express GL-coefficients in the following form:
\begin{equation}
A=A_{0}K_{A};\qquad   C=C_{0}K_{C},
\end{equation}
where $A_{0}$ and $C_{0}$ are just the usual GL-coefficients for the case of 
isotropic $s$-wave pairing \cite{PG}:
\begin{equation}
A_{0}=N(0)\frac{T-T_{c}}{T_{c}};\qquad
C_{0}=N(0)\frac{7\zeta(3)}{48\pi^{2}}\frac{v_F^2}{T_c^2},   \label{7b}
\end{equation}
where $v_{F},N(0)$-are electron velocity and normal density of 
states at the Fermi level, and everything specific to our model 
is contained in dimensionless combinations $K_{A}$ and $K_{C}$.
In the absence of impurities for both models we obtain: $K_A^0=1$, 
$K_C^0=3/2$.  For the impure system we get:

(A) $d$-wave pairing:
\begin{equation}
K_{A}= 
\frac{\gamma_0}{4\pi T_{c}}\int_{-\omega_{c}}^{\omega_{c}}\frac{d{\xi}}{\xi} 
\int_{-\infty}^{\infty}d{\omega}
\frac{\omega+\xi}{(\omega^{2}+\gamma_0^2)
ch^2\left(\frac{\omega+\xi}{2T_{c}}\right)}+
\label{8a} 
\end{equation} 
$$\frac{\gamma_1(2\gamma_0+\gamma_1)}{4T_c}\int_{-\infty}^{\infty}d{\omega}
\frac{\omega^2}{(\omega^{2}+\gamma_0^2)(\omega^2+(\gamma_0-\gamma_1)^2))
ch^2\left(\frac{\omega}{2T_{c}}\right)},$$
\begin{equation} 
K_{C}=\frac{3\pi 
T_c}{7\zeta(3)\gamma_1}\left\{\frac{2\pi T_c}{\gamma_1} 
\biggl[\Psi\left(\frac{1}{2}+\frac{\gamma_0-\gamma_1}{2\pi T_{c}}\right)
-\Psi\left(\frac{1}{2}+\frac{\gamma_0}{2\pi T_{c}}\right)
\biggr]+\Psi'\left(\frac{1}{2}+\frac{\gamma_0-\gamma_1}{2\pi T_{c}}\right) 
\right\};
\label{9a} 
\end{equation}

(B) anisotropic $s$-wave pairing:
\begin{equation}
K_{A}=\frac{\gamma_0}{\pi 
T_{c}}\Biggl\{\frac{1}{4}\int_{-\omega_{c}}^{\omega_{c}} 
\frac{d{\xi}}{\xi}\int_{-\infty}^{\infty}d{\omega}\frac{\omega+\xi}
{(\omega^2+\gamma_0^2)ch^2\left(\frac{\omega+\xi}{2T_{c}}\right)}+
\frac{2\gamma_0}{\pi}\int_{-\infty}^{\infty}d{\omega}\frac{1}
{(\omega^2+\gamma_0^2)ch^2\left(\frac{\omega}{2T_{c}}\right)}\Biggr\}, 
\label{10} 
\end{equation} 
\begin{equation} 
K_{C}=-\frac{3(\pi^2-8)}{28\pi^2\zeta(3)}\Psi''\biggl(\frac{1}{2}+
\frac{\gamma_0}{2\pi T_{c}}\biggr)
+\frac{24\pi^2}{7\zeta(3)\gamma_0^2}\frac{T_c^2}{(\pi^2-8)\gamma^2}
ln\biggl(\frac{T_{c}}{T_{c0}}\biggr)+
\frac{6\pi}{7\zeta(3)}\frac{T_{c}}{\gamma_0}.  
\label{11}
\end{equation}
The appropriate dependencies of dimensionless coefficients on disorder 
parameter $\gamma_0/T_{c0}$  in the case of $d-$wave pairing and for 
different values of the normalized anisotropic scattering rate 
$\gamma_{1}/\gamma_{0}$ are shown in Figs.3,4.

\section{Upper Critical Field}
GL-coefficients $A$ and $C$,\ as usual,\ define the temperature dependence of
the upper critical magnetic field close to $T_{c}$:
\cite{PG}:
\begin{equation}
H_{c2}=\frac{\phi_{0}}{2\pi\xi^2(T)}= -\frac{\phi_{0}}{2\pi}\frac{A}{C} 
\label{12}
\end{equation}
where $\phi_{0}=c\pi/e$ -- is magnetic flux quantum, 
$\xi(T)$ -- is temperature dependent coherence length.\ 
Now we can easily 
find the ``slope'' of the temperature dependence of $H_{c2}(T)$ near 
$T_{c}$,\ i.e.  the temperature derivative:  
\begin{equation} 
\left|\frac{dH_{c2}}{dT}\right|_{T_c}=\frac{24\pi\phi_{0}}{7\zeta(3)v_F^2}T_{c}
\frac{K_A}{K_C} \label{13b}
\end{equation}
In the case of the usual $s$-wave superconductivity  anisotropic scattering 
does not influence the behavior of the slope of the upper critical field. 
The appropriate dependencies of dimensionless parameter 
$h=|dH_{c2}/dT|_{T_c}/|dH_{c2}/dT|_{T_c0}$ on disorder $\gamma_0/T_{c0}$
in the case of $d-$wave pairing - for 
different values of the normalized anisotropic scattering rate 
$\gamma_{1}/\gamma_{0}$ are shown in Fig.5. In the case of anisotropic
$s$-wave pairing the slope as usual \cite{PS} grows with disorder and 
in the limit of strong disorder $\gamma_0\gg T_{c}$ it crosses over to the 
usual linear dependence described by the well-known  Gorkov's 
expression \cite{G}:  
\begin{equation} 
\frac{\sigma}{N(0)}\left|\frac{dH_{c2}}{dT}\right|_{T_c}
=\frac{8e^2}{\pi^2}\phi_{0}
\label{Gr}
\end{equation}  
where $\sigma=N(0)e^2v_F^2/3\gamma_0$ - is electron conductivity in the normal
state, which is characteristic of the impure superconductors with isotropic 
$s$-wave pairing. It means that strong disordering suppresses gap anisotropy 
and we obtain a usual limit of the impure superconductor.

For the case of $d$-wave pairing the slope of $H_{c2}$ drops to zero on the 
scale $\gamma_0\sim T_{c0}$ for the small values of the rate 
$\gamma_1/\gamma_0$. For the values of anisotropic scatterong rate  
on the interval $0.5\leq\gamma_1/\gamma_0\leq 0.6$ 
the behavior of the slope is qualitatively changed: $h$ smoothly but 
nonlinearly increases with the growth of $\gamma_{0}/T_{c0}$, 
crosses over the maximum and then has a sharp drop.
The interval where the slope grows extends as $\gamma_1$ approaches  
$\gamma_0$. In our opinion these sharp anomalies in dependence of the slope 
of the upper critical field on disorder can be used for determining the 
pairing type and possible role of anisotropic impurity scattering in 
"unusual" superconductors. Unfortunately, in case of high-$T_c$ oxides the 
situation is complicated by the known nonlinearity of temperature dependence 
of $H_{c2}$, which is observed in rather wide region close to $T_c$ and 
and also by some ambiguity in experimental methods to measure $H_{c2}$.
 
This work was partly supported by the grant No.96-02-16065 of the Russian 
Foundation of Fundamental Research. It was performed under the project   
IX.1 of the State Program "Statistical physics" as well as  the project 
No.96-051 of the State Program on HTSC of the Russian Ministry of Science. 

\newpage
\appendix
\section{Vertex part
$\Gamma_{{\bf pp'}}$ in ``ladder'' approximation.} 

The Bethe-Salpeter equation for the vertex part takes the form:
\begin{equation}
\Gamma_{{\bf pp'}}=U({\bf p,p'})+\sum_{{\bf p''}}U({\bf p,p''})G^R({\bf p''})
G^A({\bf p''})\Gamma_{{\bf p''p'}},
\label{A1}
\end{equation}
where $U({\bf p,p'})$-is irreducible vertex function.
We take $U({\bf p,p'})$ in the following form ("ladder" approximation):
\begin{equation}
U({\bf p,p'})=\rho V_0^2+\rho V_1^2f({\bf p})f({\bf p'}).
\label{A2}
\end{equation}
Then Eq.(A1) can be written as: 
\begin{equation}
\Gamma_{{\bf pp'}}=\rho V_0^2+\rho V_1^2f({\bf p})f({\bf p'})+
\rho V_0^2\Psi({\bf p'})+\rho V_1^2f({\bf p})\Phi({\bf p'})
\label{A3}
\end{equation}
where
\begin{eqnarray}
\Psi({\bf p'})=\sum_{{\bf p''}}G^R({\bf p''})G^A({\bf p''})
\Gamma_{{\bf p''p'}},\\ 
\label{A4}
\Phi({\bf p'})=\sum_{{\bf p''}}f({\bf p''})G^R({\bf p''})G^A({\bf p''})
\Gamma_{{\bf p''p'}}. \nonumber
\end{eqnarray}
From Eq.(A3) one can obtain a self-consistent set of equations for 
$\Psi({\bf p'})$ and $\Phi({\bf p'})$:
\begin{equation} 
\left\{\begin{array}{ll} 
\Psi({\bf p'})=\rho V_0^2I_1+\rho V_1^2f({\bf p'})I_2+
\rho V_0^2I_1\Psi({\bf p'})+\rho V_1^2I_2\Phi({\bf p'}),
\\ \Phi({\bf p'})=\rho V_0^2I_2+\rho V_1^2f({\bf p'})I_3+
\rho V_0^2I_2\Psi({\bf p'})+\rho V_1^2I_3\Phi({\bf p'}),
\end{array} \right.  
\label{A5} 
\end{equation}
where
\begin{eqnarray}
I_1=\sum_{{\bf p}}G^R({\bf p})G^A({\bf p}), \nonumber \\
I_2=\sum_{{\bf p}}f({\bf p})G^R({\bf p})G^A({\bf p}), \label{A6} \\
I_3=\sum_{{\bf p}}f^2({\bf p})G^R({\bf p})G^A({\bf p}). \nonumber
\end{eqnarray}
Solving system (A5), one can find the appropriate expressions for 
$\Psi({\bf p'})$ and $\Phi({\bf p'})$ and hence the expression for the vertex 
part:
\begin{equation}
\Gamma_{{\bf pp'}}=\frac{\rho V_0^2(1-\rho V_1^2I_3+
\rho V_1^2f({\bf p'})I_2)+\rho V_1^2(f({\bf p})f({\bf p'})(1-\rho V_0^2I_1)
+\rho V_0^2f({\bf p})I_2)}{(1-\rho V_0^2I_1)(1-\rho V_1^2I_3)
-\rho V_0^2\rho V_1^2I_2^2} 
\label{A7} 
\end{equation}

\newpage
\section{Ginzburg-Landau coefficients.}
We can easily see that the contribution of the diagram Fig.2(a) is
\begin{eqnarray}
-\frac{T}{(2\pi)^2}\Delta_q^2\sum_{\omega}\int{d{\bf p}}2cos^2(2\phi)
G_{\omega}({\bf p_{+}})G_{-\omega}({\bf p_{-}})= \nonumber \\
-\Delta_q^2TN(0)\sum_{\omega}\int \frac{d\xi}{\tilde\omega^2+\xi^2}+
\Delta_q^2q^2\frac{N(0)\pi v_{F}^2T_{c}}{8}\sum_{\omega}
\frac{1}{|\tilde\omega|^3}.
\label{B1}
\end{eqnarray}
The contribution of the diagram Fig.2(b) is
\begin{equation}
-\frac{T}{(2\pi)^2}\Delta_q^2\sum_{\omega}\int{d{\bf p}}2cos^2(2\phi)
G_{\omega}({\bf p})G_{-\omega}({\bf p})= 
-\Delta_q^2T_{c}N(0)\sum_{\omega}\int \frac{d\xi}{\tilde\omega^2+\xi^2}.
\label{B2}
\end{equation}
The contribution of the diagram with diffusion propagator Fig.2(b) is
\begin{equation}
-T\sum_{\omega}\sum_{\bf pp'}\sqrt{2}cos(2\phi)G^R({\bf p_{+}})
G^A({\bf p_{-}})\Gamma_{{\bf pp'}}\sqrt{2}cos(2\phi')G^R({\bf p'_{+}})
G^A({\bf p'_{-}}).
\label{B3}
\end{equation}
Taking into account (A6) and (A7) we get from here
\begin{equation}
-TN(0)\pi\gamma_{1}\sum_{\omega}\left[\frac{1}
{|\tilde\omega|(|\tilde\omega|-\gamma_1)}-
\frac{v_F^2(2|\tilde\omega|-\gamma_1)q^2}{8|\tilde\omega|^3
(|\tilde\omega|-\gamma_1)^2}\right]. 
\label{B4} 
\end{equation}
Note that in the absence of anisotropic scattering for the 
case of $d-$pairing the contribution of diagrams (c) actually 
vanishes up to terms of the order of $q^2$.

In the same way we get the appropriate contribution of the diagram (d)
\begin{equation} -T_áN(0)\pi\gamma_{1}\sum_{\omega}\frac{1} 
{|\tilde\omega|(|\tilde\omega|-\gamma_1)}.
\label{B5} 
\end{equation}
Finally we get the expression for $F_s-F_n$ and so 
Ginzburg-Landau coefficients cited in the main body of the paper.

\newpage
\begin{center}
{\bf FIGURE CAPTIONS}
\end{center}

Fig.1. Critical temperature $T_{c}$ as a function of the normalized isotropic 
scattering rate $\gamma_0/T_{c0}$. The dashed curve represents the $s$-wave 
pairing case, the solid curves represent the $d$-wave pairing case for 
different values of the normalized anisotropic scattering rate  
$\gamma_1/\gamma_0$:  

1---$\gamma_1/\gamma_0$
=0.0; 2---0.3; 3---0.5; 4---0.6; 5---0.7; 6---0.8; 7---0.9; 8---0.95. 

Fig.2. Diagrammatic representation of Ginzburg-Landau expansion. Electronic 
lines are "dressed" by impurity scattering. $\Gamma$ is the impurity vertex 
calculated in "ladder" approximation. Diagrams (c) and (d) are calculated 
with $q=0$ and $T=T_{c}$.

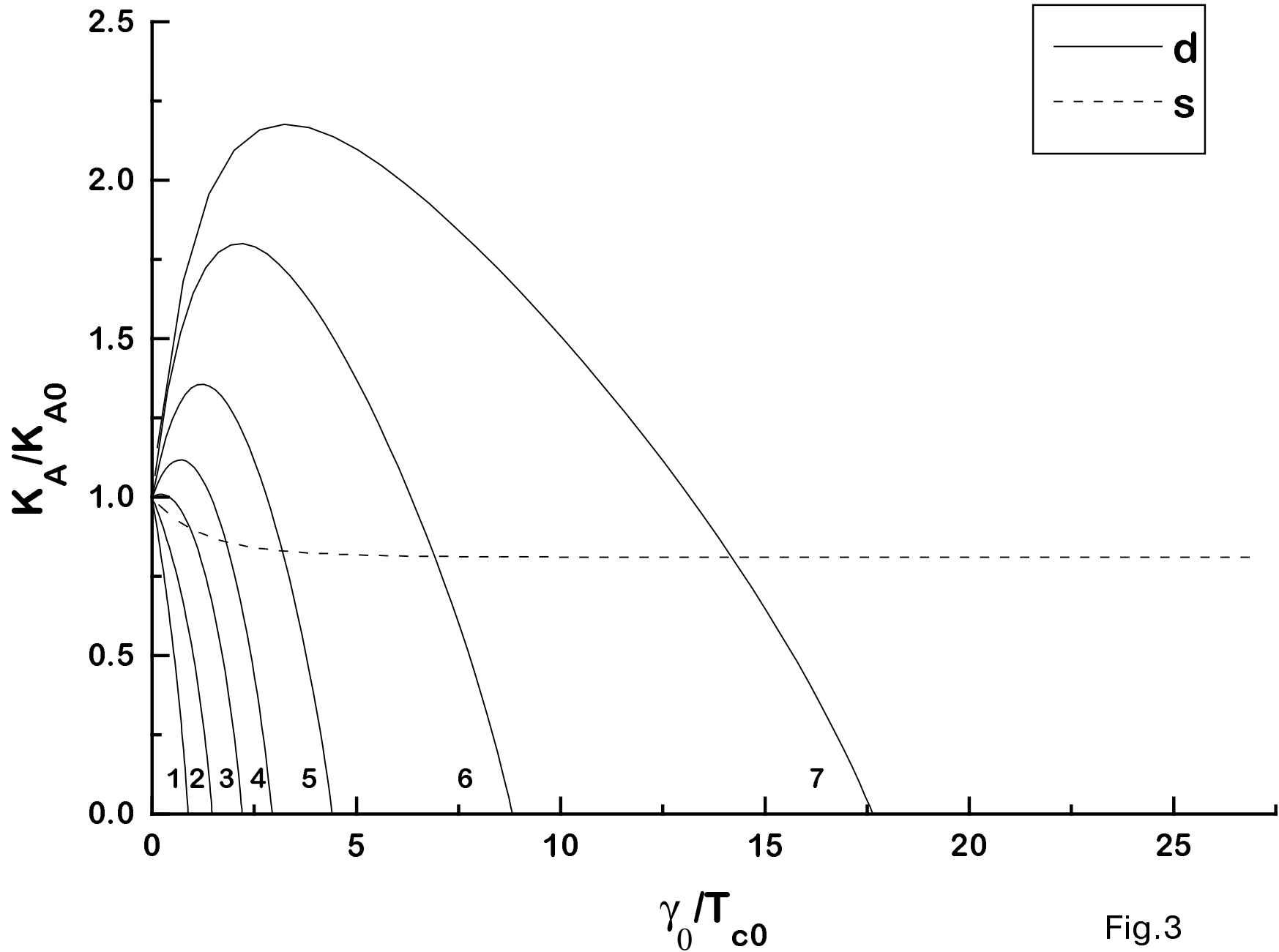
Fig.3. Dependence of dimensionless coefficient $K_A/K_{A0}$ on disorder 
parameter $\gamma_0/T_{c0}$. The dashed curve represents the $s$-wave 
pairing case, the solid curves represent the $d$-wave pairing case for 
different values of the normalized anisotropic scattering rate  
$\gamma_1/\gamma_0$:  

1---$\gamma_1/\gamma_0$
=0.0; 2---0.4; 3---0.6; 4---0.7; 5---0.8; 6---0.9; 7---0.95.

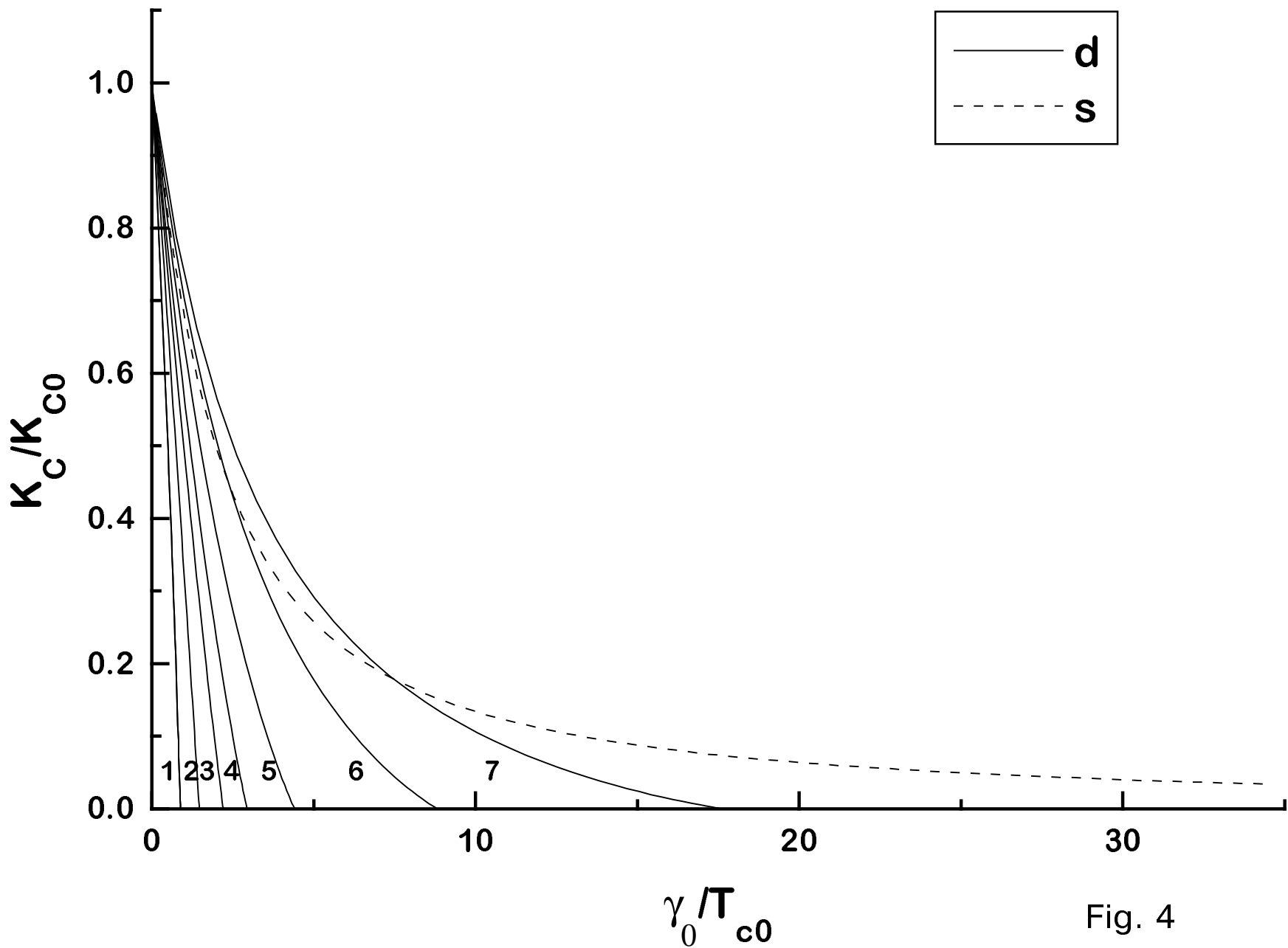
Fig.4. Dependence of dimensionless coefficient $K_C/K_{C0}$ on disorder 
parameter $\gamma_0/T_{c0}$. The dashed curve represents the $s$-wave 
pairing case, the solid curves represent the $d$-wave pairing case for 
different values of the normalized anisotropic scattering rate  
$\gamma_1/\gamma_0$:  

1---$\gamma_1/\gamma_0$
=0.0; 2---0.4; 3---0.6; 4---0.7; 5---0.8; 6---0.9; 7---0.95.

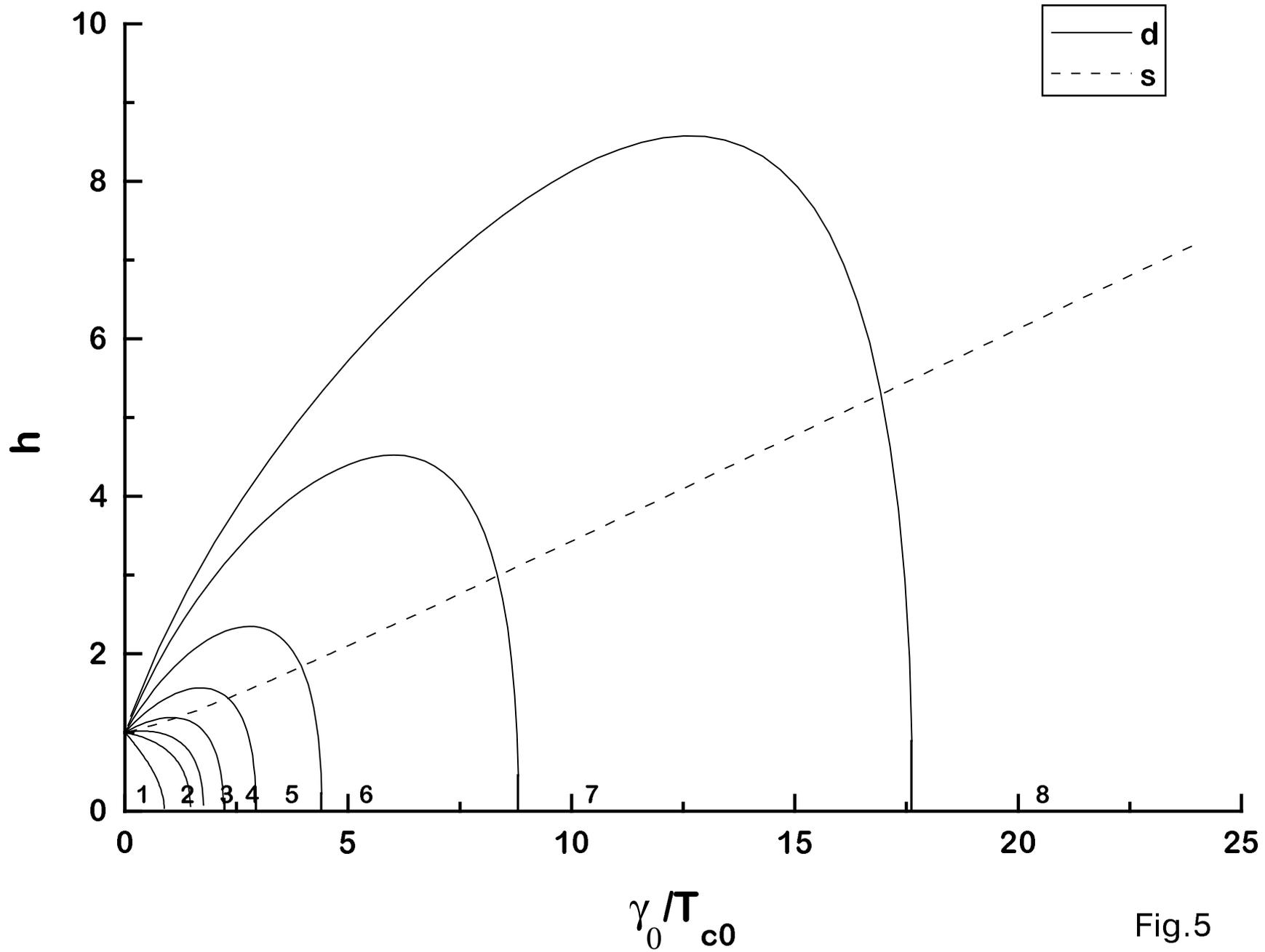
Fig.5. Dependence of normalized slope of the upper critical field
$h=\left|\frac{dH_{c2}}{dT}\right|_{T_c}/
\left|\frac{dH_{c2}}{dT}\right|_{T_{c0}}$  on disorder 
parameter $\gamma_0/T_{c0}$. The dashed curve represents the $s$-wave 
pairing case, the solid curves represent the $d$-wave pairing case for 
different values of the normalized anisotropic scattering rate  
$\gamma_1/\gamma_0$:  

1---$\gamma_1/\gamma_0$
=0.0; 2---0.4; 3---0.5; 4---0.6; 5---0.7; 6---0.8; 7---0.9; 8---0.95.

\end{document}